# Optimal Control of a Single Queue with Retransmissions: Delay–Dropping Tradeoffs

Anastasios Giovanidis, *Student Member, IEEE,* Gerhard Wunder, *Member, IEEE,* Jörg Bühler, *Student Member, IEEE*

**Abstract**

A single queue incorporating a retransmission protocol is investigated, assuming that the sequence of per effort success probabilities in the Automatic Retransmission reQuest (ARQ) chain is a priori defined and no channel state information at the transmitter is available. A Markov Decision Problem with an average cost criterion is formulated where the possible actions are to either continue the retransmission process of an erroneous packet at the next time slot or to drop the packet and move on to the next packet awaiting for transmission. The cost per slot is a linear combination of the current queue length and a penalty term in case dropping is chosen as action. The investigation seeks policies that provide the best possible average packet delay-dropping trade-off for Quality of Service guarantees. An optimal deterministic stationary policy is shown to exist, several structural properties of which are obtained. Based on that, a class of suboptimal $< L, K >$-policies is introduced. These suggest that it is almost optimal to use a K-truncated ARQ protocol as long as the queue length is lower than L, else send all packets in one shot. The work concludes with an evaluation of the optimal delay-dropping tradeoff using dynamic programming and a comparison between the optimal and suboptimal policies.

[0]This work is supported by the German Federal Ministry of Education and Research as part of the ScaleNet project 01BU566.

The authors are with the Fraunhofer German-Sino Mobile Communications Lab, Heinrich-Hertz-Institut, Einsteinufer 37, D-10587 Berlin, Germany. Tel/Fax: +493031002-860/-863, e-mail: {giovanidis, wunder, buehler} @ hhi.de and with the Technical University of Berlin, Heinrich Hertz Chair for Mobile Communications, Werner-von-Siemens-Bau (HFT 6), Einsteinufer 25, D-10587 Berlin, Germany.

# I. Introduction

Retransmission protocols are applied in communications over fading channels to achieve reliability. The concept of such protocols is to detect erroneous packets at the receiver and then request retransmission for these data. The information whether the message has been detected as correct or erroneous is sent back to the transmitter via a binary feedback link using the signal NACK to request a new retransmission or ACK to declare that the packet is correctly received. In the latter case, the sender moves on to the first transmission of the next packet waiting in the buffer. In this way error-free communications can be guaranteed. However, reliability naturally comes at a cost. The costs include a waste of resources, e.g. power or spectrum, for the unsuccessful efforts, a reduction in transmission rate when error correction and detection parity bits are added to the original code, as well as a noteworthy increase in packet delay due to multiple channel reuses for the correct transmission of a single message. During the last decades, different types of retransmission protocols have been proposed in the literature (see [1], [2] and references therein) and adopted by standards of mobile networks such as UMTS, WiMAX and 3GPP LTE. These include the simple Stop-and-Wait ARQ (SW-ARQ) protocol, as well as Type-I or Type-II Hybrid-ARQ (HARQ) schemes, where Forward Error Correction codes (FEC) are used together with packet combining to enhance the protocols' performance, [3], [4], [5], [6], [7].

Several efforts to optimize ARQ protocols, in the sense of reducing necessary retransmission efforts with economy in resources at the same time, can be found in the recent literature. Power control per retransmission to maximize the throughput of ARQ protocols has been investigated in [8]. In [9] the optimal sequence of redundancy per effort is found based on a dynamic programming formulation. The work in [10] investigates the optimal power and rate allocation among SW-ARQ retransmissions, that guarantees average delay or throughput constraints, based on some channel state information (CSI) at the receiver. A trade-off between throughput and energy consumption when the CSI is partially observable is given in [11] through a semi-Markov Decision Process formulation, while delay- and overflow-aware joint rate and power adaptation for type-I and power adaptation for type-II HARQ protocols is provided in [12] and [13], respectively. Rather noteworthy is also the work in [14], where the average delay of a single user wireless communication system, which includes a buffer and incorporates SW-ARQ retransmissions, is optimized by combined power and rate control under average power constraints. Furthermore, in [15] optimal stopping arguments have been used to determine the optimal maximum retransmission number, with respect to a cost function which lineary combines throughput gain with delay and packet dropping costs.

In the current work, a single queue incorporating an ARQ protocol is investigated. The per effort sequence of success probabilities $\{q_k\}$ in each ARQ round, up to correct packet reception, is a priori defined, the same approach found also in [16]. The motivation behind this is that the success probabilities generally depend on the amount of parity bits, type of modulation, transmission power and type of protocol used. When no channel state information is available, an optimal fixed choice of $\{q_k\}$ can be calculated offline (see also the investigations in [15] and [17]). Using a 2-dimensional state-space for the system (current queue length and retransmission number), we formulate a Markov Decision Problem (MDP) with average cost criterion [18] where at each state we have to choose between *continuing* retransmitting the erroneous packet and *dropping* it to proceed to the first transmission of the next packet waiting. The investigation here seeks optimal policies that provide the best possible tradeoff between average delay and percentage of dropped packets. These are two conflicting performance mesures both of great importance concerning Quality of Service (QoS) guarantees in communications.

The model under study as well as the MDP formulation are presented in Sections II and III. Several structural properties of the optimal policy are proved in Section IV. Section V introduces a family of sub-optimal policies, named here $< L, K >$*-policies*, which encapsulate the structural properties of section IV and at the same time simplify the design considerably. The idea is that it is almost optimal to choose a truncated ARQ protocol having $K$ as fixed maximum number of retransmissions and use it for reliable communications, as long as the queue length does not exceed a certain threshold L, after which the packets are sent in one shot and no more retransmissions take place. For the optimal policies on the other hand, the number of maximum allowable retransmissions is not constant but varies depending on the queue length. Evaluation of *the optimal delay-dropping tradeoff* and comparison between the optimal policy and the sub-optimal $< L, K >$-policies is presented in Section VI. Section VII concludes our work. Proofs of theorems from the main text are provided in the appendix.

## II. A Single Queue with Retransmissions

Consider a single-server communication system, where data packets arriving from some information source are stored in a buffer while waiting for service. The time axis is considered time-slotted, all slots having equal length $T$. During each slot an arrival of a new packet may occur (or not) with probability $0 < \lambda \leq 1$ (respectively $1 - \lambda$), i.e. the arrival processes is Bernoulli. The service rate is considered constant $\mu = 1$ $[packet/slot]$ and a transmission of a packet is *attempted* at the end of each time slot under the condition that the queue is not empty.

Due to the stochastic nature of the wireless channel, communications are generally unreliable and errors may occur. We assume that with use of error detection codes, erroneous packets are detected at the receiver with probability $1$. The transmitter is informed via a zero-delay error-free feedback link whether decoding of the packet has been successful or not. A sequence of random variables $X_n \in \{0, 1\}\,,\ n \in \mathbf{N}$ is used for the information fed back after decoding of a packet at slot $n$. $X_n = 0$ denotes no acknowledgement (NACK) whereas $X_n = 1$ acknowledgement (ACK). It is noted here that in real systems, although the feedback link can be made rather reliable, the delay is often considerable. The current work does not address issues on delayed feedback to keep simplicity of the model. In case an ACK is fed back the packet is removed from the buffer and the first transmission of the next packet waiting in queue takes place during the next time slot (a nonpreemptive First Come First Serve service principle is assumed). On the other hand in case a NACK is fed back there exist two possibilities. Either the transmission of the erroneous packet is repeated at the next slot or the packet is discarded (we say that *dropping* occurs) and the next waiting packet is served during the $n+1$-th slot. The two posibilities may be considered as *actions* that a controller may choose in order to optimize the system performance. For each slot we introduce further the pair of random variables (r.v.'s) $(U_n, Z_n)$. The first one provides the current queue length whereas the second one the current retransmission number at slot $n$. The pair takes values within the set $\mathbf{N} \times \mathbf{N}_+$, $\mathbf{N}$ and $\mathbf{N}_+$ being the set of non-negative and strictly positive integers respectively. As explained in the introduction there exist various types of Automatic Retransmission reQuest (ARQ) protocols used in practical communications systems. In this work we describe each ARQ protocol by the one-step transition probability matrix [15]

$$\mathbf{P}_{ARQ} = \begin{pmatrix} q_1 & p_1 & 0 & 0 & 0 & \cdots \\ q_2 & 0 & p_2 & 0 & 0 & \cdots \\ q_3 & 0 & 0 & p_3 & 0 & \cdots \\ \vdots & \vdots & \vdots & \vdots & \ddots & \vdots \end{pmatrix} \tag{1}$$

with unequal entries in general. The above matrix specifically describes the random process $\{Z_n\}$, where as mentioned before $Z_n$ is the current number of retransmissions at time slot $n$ and refers to non-truncated protocols which repeat transmissions as many times as necessary up to correct reception. It is typical of a specific type of random walk in one dimension better known as a *success run*. We denote as a *retransmission policy* a specific choice of the sequence of success probabilities $\{q_k\}$.

The random variables $X_n$ and $Z_n$ are dependent. This is formally described as follows. The conditional probability of successful reception at stage $k$ equals $\mathbf{Pr}\left(Z_{n+1} = 1 | Z_n = k\right) = \mathbf{Pr}\left(X_n = 1 | Z_n = k\right) = q_k$ and of failure $\mathbf{Pr}\left(Z_{n+1} = k+1 | Z_n = k\right) = \mathbf{Pr}\left(X_n = 0 | Z_n = k\right) = p_k$. Obviously $\{Z_n\}$ forms a Markov chain. On the other hand the sequence $\{X_n\}$ depends on the number of time slots since the last ACK was received and is thus history dependent. Since the events are mutually exclusive and exhaustive it holds $p_k + q_k = 1$, $\forall k = 1, 2, \ldots$ and the matrix in (1) is stochastic. Conditions for its ergodicity can be found in [15]. The transition probability diagram is given in fig. 2. In the following the notation $p\left(k\right)$ is often used instead of $p_k$ and $p\left(Z_n\right)$ instead of $p_{Z_n}$ for simplicity.

The matrix in (1) can be used as a general description of different types of ARQ protocols if no CSI (accurate or estimated) is available at the transmitter at any time, hence no link adaptation to current channel conditions is performed. This is standard in the literature (see [19], [20] and [2]). ACK/NACK is the only feedback assumed, possibly combined with information about the channel fading statistics. The latter can be infrequently sent to the transmitter. The values of the sequence $\{q_k\}$ generally depend on the available knowledge about the channel, the resources allocated per trial $k$, the applied modulation and coding schemes, as well as the type of the protocol used, which may exploit - or not - information from previous erroneous efforts. For Stop-and-Wait protocols errors occur in the case of channel outages and the success probability function depends only on the allocated power and coding rate of the current transmission. An optimization problem formulated in [17] finds the optimal power allocation for each retransmission effort and hence the optimal *retransmission policy*, given delay and dropped packet penalties. In the case of HARQ protocols CSI is also not necessary. Upon detection of a transmission failure, a NACK signal is fed back and only redundant information is retransmitted. The receiver combines the soft information of original transmission with subsequent retransmissions to achieve a higher probability of successful decoding and also improve throughput performance. The sequence of success probabilities then depends on the way the soft combining is performed. In the case of Chase Combining the same packet is repeatedly sent and the receiver aggregates the retransmission energy. If Incremental Redundancy (IR) is applied redundancy bits are produced by using e.g. Rate Compatible Punctured Convolutional Codes (RCPC) and Turbo-codes. The redundancy bits are sent only when an error occurs and combined with the erroneous packets at the receiver increase the probability of correct reception step-wise. In [2] closed-form expressions for the success probabilities regarding the aforementioned protocols can be found, whereas for the case of HARQ with IR, the optimal partitioning of parity bits among retransmissions is found in [21]. In [9] the optimal Type II HARQ *retransmission policy* is found using dynamic programming. In all examples mentioned above, as well as in [16], the *retransmission policy* is fixed and does not vary adaptively with the channel conditions.

## III. A Discrete Time Markov Decision Problem with Average Cost Criterion

In the following we describe the decision problem under study. Consider a Markov Decison Process [18] $\{\mathcal{T}, \mathcal{S}, \mathcal{A}, \mathbf{Pr}\left(\bullet|S,A\right), c\left(S,A\right)\}$, with set of decision epochs $\mathcal{T} = \{0,1,\ldots\}$ and state space $\mathcal{S} = \mathbf{N}\times\mathbf{N}_+$ the state in our problem being the pair of random variables $S_n = (U_n, Z_n)$ at $n$. The action set is binary $\mathcal{A} = \{0,1\}$, $\mathbf{Pr}\left(\bullet|S,A\right)$ is the transition probability distribution of the system which is conditioned on the current state and action, whereas $c$ is the system cost per time slot as a function of state $S_n$ and action $A_n$, $c : \mathbf{N}\times\mathbf{N}_+\times\{0,1\}\rightarrow\mathbf{R}_+$.

At the beginning of slot $n$ the state information $S_n = (U_n, Z_n)$ is available at the controller. Using possibly some further knowledge over the entire history $h_n = \{S_1, A_1, \ldots, A_{n-1}\}$ of previous states and actions taken the controller may choose between two actions, which will affect the packet transmission at the *next slot* $n+1$. Either choose action $A_n = 0$ and continue the retransmission cycle in the next slot in case a NACK is fed back ($X_n = 0$), or choose $A_n = 1$ and break the retransmission cycle irrespective of the result of decoding of the transmitted packet in $n$ ($X_n \in \{0,1\}$) and begin in $n+1$ a new transmission round. It is important to note that choosing action $A_n = 1$ does not necessarily result in a dropped packet. The packet will be actually dropped only in the case of $X_n = 0$, which will occure with probability $p\left(Z_n\right)$.

During slot $n$ an arrival of a new packet $\alpha_n \in \{0,1\}$ may occur with probability $\lambda$, which will be taken into account for the value of the queue length at the next state, together with the result of the current packet transmission $X_n$. Then for the state evolution $S_{n+1} = (U_{n+1}, Z_{n+1})$

$$U_{n+1} = \begin{cases} [U_n - X_n]^+ + \alpha_n & \text{if } A_n = 0 \\ [U_n - 1]^+ + \alpha_n & \text{if } A_n = 1 \end{cases} \tag{2}$$

$$Z_{n+1} = \begin{cases} 1 + Z_n (1 - X_n) & \text{if } A_n = 0 \\ 1 & \text{if } A_n = 1 \end{cases} \tag{3}$$

The state transition probabilities are given in table I.

In the scenario described above two measures are rather important for the performance of the single queue system. The first one is the average *queue length* $\bar{U}$, related by Little's Law $\bar{U}/\lambda = \bar{W}$ [22] to the average packet delay (average waiting time per packet in the buffer) $\bar{W}$. The second is the average number of dropped packets $\bar{\Delta}$ in case the ARQ cycle is interrupted by the controller prior to correct packet reception. A cost per time slot is introduced which equals the actual queue length while dropping is incorporated as a penalty, weighted by $\delta > 0$. As mentioned earlier, dropping occurs if action $A_n = 1$ is chosen at slot $n$ *and* the packet is not correctly received ($X_n = 0$), i.e. the dropping cost term equals $\delta A_n (1 - X_n)$. Since the result of decoding is known at the end of each slot and the cost per slot depends on the random variable $X_n$, we use an expected cost, the expectation taken over the random disturbance $X_n$ (see [18, p. 20]). In this way the cost can be written as a function of the state and action taken

$$c_\delta \left( (U_n, Z_n), A_n \right) = \mathbf{E}_{X_n} \left[ U_n + \delta A_n (1 - X_n) \right] = U_n + \delta A_n p(Z_n) \tag{4}$$

For the case where $U_n = 0$ we have $c_\delta \left( (0, Z_n), A_n \right) = 0$. In this work we aim to find a *deterministic stationary policy*, which is optimal among all history dependent randomized policies $\pi \in \Pi^{HR}$, in terms of minimizing the average expected cost, for a certain value of the weight $\delta > 0$ and initial state $S_o$

$$J_\pi (\delta, S_o) := \limsup_{N \to \infty} \frac{1}{N} \mathbf{E}^{\pi, S_o} \left[ \sum_{n=1}^{N} c_\delta (S_n, A_n) \right] \tag{5}$$

The optimal strategy $\pi^* \in \Pi^{HR}$ is defined as the one that satisfies $J_{\pi^*}(\delta, S_o) = J^*(\delta, S_o) \leq J_\pi (\delta, S_o), \ \forall \pi \in \Pi^{HR}$. Since the model can be easily verified to be *unichain* [18, pp. 348-354] (the transition matrix of each stationary deterministic policy has a single positive recurrence class) the optimal average cost (if it exists) is equal for all initial states $S_o$.

The $\limsup$ average cost may not be finite $\forall \pi \in \Pi^{HR}$. This situation occurs in two cases. (a) The chain in (1) is non-ergodic (see [15]) and hence there exists a positive probability for the ARQ chain never to return to state $Z_n = 1$, or (b) the arrival rate is greater than the minimum average service rate $\lambda > \mu_{\min} := \left(1 + \sum_{k=1}^{\infty} p_1 \ldots p_k\right)^{-1}$ [15], [23], the latter defined as the inverse of the expected retransmission number, in case no dropping occurs. In such cases the queue is unstable and all queue states are recurrent-null or transient. However there definitely exist policies for which the average cost remains finite. Two such families important for the investigation are provided in the following.

**Definition 1** *A stationary policy truncating the ARQ protocol up to the $K$-th retransmission number for all states of the queue length so that $A_n = 1$ if $Z_n \geq K$ is called a **K-retransmitting policy**.*

**Definition 2** *A policy which does not allow retransmissions after a certain queue length $L$, in other words $A_n = 1$, if $U_n \geq L$ is called a **L-truncating policy**.*

These two families definitely stabilize the single queue system. To see this, by appropriate choice of $K$ in the first case we can have $\hat{\mu} = \left(1 + \sum_{k=1}^{K-1} p_1 \ldots p_k\right)^{-1} \geq \lambda > \mu_{\min}$. For the second case, since the arrival process is Bernoulli, the queue length remains upper bounded by $L$ (see (2) for $A_n = 1$ and $U_n = L$) for all $n$.

### *A. Vanishing Discount Approach*

The solution of Markov Decision Problems with an average expected cost criterion is often studied in the literature [18], [24], [25] as the limit behavior of $\beta$-discounted models when the discount factor $0 < \beta < 1$ tends to 1. The *discounted expected cost* related to (5) is given by

$$J_{\pi,\beta}(\delta, S_o) := \lim_{N\to\infty} \mathbf{E}^{\pi,S_o}\left[\sum_{n=1}^{N} \beta^{n-1} c_\delta(S_n, A_n)\right] \tag{6}$$

$J_\beta^*(\delta, S_o) \leq J_{\pi,\beta}(\delta, S_o), \forall \pi \in \Pi^{HR}$ is the solution of the discounted minimization problem. We will from now on neglect the dependence on $\delta$ in notation. From [18, Th.8.10.7 and 8.10.9] we have that under mild assumptions which can be verified for the problem at hand

$$J^* = \lim_{\beta\to 1}(1-\beta) J_\beta^*(S_o) \tag{7}$$

for all $S_o \in \mathcal{S}$. Furthermore, from [24, Th.3.8] the optimal policy $\pi^*$ is also *limiting discount optimal* in the sense that there exists a sequence $\beta_m \to 1$ and $S_m \to S$ such that $\pi^*(S) = \lim_{m\to\infty} \pi_{\beta_m}(S_m)$, $\forall S \in \mathcal{S}$. We may thus focus on discounted cost optimality and get the solution to the average cost problem passing to the limit $\beta \to 1$ in (7). See also the works [26] and [11].

For the discounted expected cost problem we provide the Bellman optimality equations of the problem at hand, for each state $S := (U, Z) = (l, k) \in \mathcal{S}$. These are written as $J_\beta^*(l,k) = TJ_\beta^*(l,k)$ [18, pp. 146-148], where

$$\begin{aligned}
\mathbf{l \geq 1} \quad : \quad & TJ_\beta^*(l,k) = \min\Big\{l + \beta\Big[q_k(1-\lambda) J_\beta^*(l-1,1) + q_k\lambda J_\beta^*(l,1) + \\
& +p_k(1-\lambda) J_\beta^*(l,k+1) + p_k\lambda J_\beta^*(l+1,k+1)\Big], \\
& l + \delta p_k + \beta\Big[\lambda J_\beta^*(l,1) + (1-\lambda) J_\beta^*(l-1,1)\Big]\Big\} \\
\mathbf{l = 0} \quad : \quad & TJ_\beta^*(0,k) = \beta\Big[\lambda J_\beta^*(1,k) + (1-\lambda) J_\beta^*(0,k)\Big]
\end{aligned} \tag{8}$$

The existence of an optimal solution to the above discounted optimality equations is guaranteed by the Banach fixed-point theorem [18, Th.6.2.3] and some further technical conditions [18, Th.6.10.4, Prop.6.10.5] due to the unboundedness of the costs, which can be easily shown to be satisfied for the problem at hand.

### *B. Algorithmic Solution for Finite State Space: Value Iteration*

Using value iteration algorithms [18, pp.160-161, pp. 364-365] we may find deterministic $\epsilon$-optimal policies $\pi_\epsilon$ for both discounted and average cost problems. The algorithms may be implemented *only for finite state spaces*. We argue here that bounding the maximum queue length by $L_{\max} > 0$ and the maximum allowable retransmission number by $K_{\max} > 0$ provides an approximation to the optimal solution which is improving as $K_{\max}$ and $L_{\max}$ increase. In the following we provide two examples (see Table II) of average cost optimal policies for two different sets of success probabilities $\{q_k\}, k = 1, \ldots, K$. The discount factor is $\beta = 0.99$ and the Bernoulli arrival rate is $\lambda = 0.4$ [packet/sec]. A maximum number of $K_{\max} = 6$ retransmissions is allowed, the queue length is restricted to a length of $L_{\max} = 10$ packets and the packets are *discarded* ($A = 1$) if either $K_{\max}$ or $L_{\max}$ is exceeded. In the first example a *monotone decreasing* sequence of success probabilities is utilized, specifically having values $q_k = e^{-0.9k},\ k = 1, \ldots, 6$ and weight $\delta = 40$. In the second one the sequence of success probabilities is *monotone increasing*, $q_k = 1 - e^{-0.9k},\ k = 1, \ldots, 6$ and $\delta = 4$. The examples above suggest that the optimal policy for the actual problem behaves monotonically in both the $l$- (queue length-) and $k$- (number of efforts-) axis. Observe that the 1's in the last column which imply dropping and break the monotonicity in the second example are simply a result of the finite state-space and will not appear in the actual problem with countably infinite state space.

## IV. Structural Properties of Optimal Policies

As the examples in Table II indicate, for certain senarios, if it is optimal to drop at some specific queue length and number of retransmissions, it is as well optimal to drop for all greater queue states for the same $k$. The same result may be found for a fixed queue length and varying the retransmission number. The analysis that follows will be based on the assumption that the sequence of conditional success probabilities $\{q_k\}$ is either monotone non-increasing or non-decreasing w.r.t. $k$. This assumption is supported by results in [17] for the optimal choice of conditional success probabilities in the SW-ARQ case as well as by the fact that $\{q_k\}$ is definitely non-decreasing in the case of Type-II HARQ.

All proofs of the following Lemmata and Theorems can be found in the Appendix. For the proofs value and policy iteration methods [18] are used. Let us first provide some monotonicity properties of the value function.

**Lemma 1** *For all states* $(l, k) \in \mathcal{S}$*,* $m \in \mathbf{N}$ *and iteration steps* $n \in \{0, 1, \ldots\}$

$$J^n(l, k) \leq J^n(l + m, k) \tag{9}$$

**Lemma 2** *For all states* $(l, k) \in \mathcal{S}$*,* $m \in \mathbf{N}$*, iteration steps* $n \in \{0, 1, \ldots\}$ *and monotone non-increasing conditional success probabilities* $q_1 \geq q_2 \geq \ldots$*, it holds*

$$J^n(l, k) \leq J^n(l, k + m) \tag{10}$$

Combining the above two Lemmata we obtain

**Corollary 1** *For all states* $(l,k) \in \mathcal{S}$*,* $n \in \{0,1,\ldots\}$ *and non-increasing success probabilities*

$$J^n(l,k) \overset{(9)}{\geq} J^n(l-1,k) \overset{(10)}{\geq} J^n(l-1,1) \tag{11}$$

We may now state the following Theorem.

**Theorem 1** *Suppose* $q_1 \geq q_2 \geq \ldots$ *Given a fixed queue state* $l$ *and varying the retransmission number* $k$*, the optimal policy is of threshold type, i.e. there exists a critical state* $\left(l, \hat{k}_l\right)$*, possibly dependent on* $l$*, such that dropping is optimal for* $k \geq \hat{k}_l$*, while continue is optimal for* $k < \hat{k}_l$*.*

A proof on the monotone behavior of the optimal policy at the l-axis could not be attained, although all examples tested suggest that, given $k$, there exists some threshold queue length $\hat{l}_k$ such that dropping is optimal $\forall l \geq \hat{l}_k$ and continuing retransmissions is optimal for $l < \hat{l}_k$. We can prove the following important property instead

**Theorem 2** *For the optimal policy* $\pi^*$ *there exists a threshold queue length* $l_{\pi^*}^{th}$ *such that* $\forall l \geq l_{\pi^*}^{th}$ *,*$\forall k$*, it holds that* $\pi^*(l,k) = 1$*. Furthermore, the threshold in the* $l$*-axis is always finite and more specifically*

$$l_{\pi^*}^{th} \leq \delta \max_k q_{k+1} + (1-\lambda) \tag{12}$$

**Theorem 3** *It is always optimal to drop - in other words* $l_{\pi^*}^{th} = 0$*, if*

$$\delta \leq \frac{1 + \lambda/(1-\lambda) + \lambda}{1 + (1-\lambda)p_1 - \min_{k \neq 1} p_k} \tag{13}$$

*The bound is non-decreasing with* $\lambda$ *and tends to* $\infty$ *for* $\lambda \to 1$*.*

The results for the existence of a threshold in the $l$-axis in Theorem 2 and the optimality of the *always-drop* policy in Theorem 3 hold irrespective of the choice of success probabilities. For the case of monotone non-decreasing success probabilities we can prove by induction using Policy Iteration equivalently to Theorem 1 that dropping is optimal for $k < \hat{k}_l$ and continue for $k \geq \hat{k}_l$, where the k-axis threshold is $l$-dependent.

**Theorem 4** *Consider the case of monotone non-decreasing success probabilities* $q_1 \leq q_2 \leq \ldots$*, with* $q_k \to 1$*, as* $k \to \infty$*. For all states* $(l,k) \in \mathcal{S}$*,* $m \in \mathbf{Z}_+$ *the following two inequalities hold for the optimal policy* $\pi^*$

$$J_{\pi^*}(l,k) \quad \geq \quad J_{\pi^*}(l-1,1) \tag{14}$$

$$J_{\pi^*}(l,k) \quad \geq \quad J_{\pi^*}(l,k+m) \tag{15}$$

*Furthermore, given a fixed queue state* $l$ *and varying the retransmission number* $k$*, the optimal policy is of threshold type, i.e. there exists a critical state* $\left(l, \hat{k}_l\right)$*, possibly dependent on* $l$*, such that dropping is optimal for* $k < \hat{k}_l$*, while continue is optimal for* $k \geq \hat{k}_l$*.*

## V. Design Rules for the Optimal Tradeoff

The above theorems provide important structural properties of the optimal policy. We may identify two very important parameters that define the structure, namely the sequence of success probabilities per retransmission $\{q_n\}$ as well as the weighting factor $\delta > 0$, which plays the role of the penalty when dropping occurs. Especially $\delta$ has a crucial role in the upper bound for $l_{\pi^*}^{th}$ provided. Specifically, Theorem 2 suggests that it is optimal to use an ARQ protocol only up to a finite queue length $l_{\pi^*}^{th}$. When the queue exceeds the threshold the packet is removed from the queue after the first effort regardless of the result of decoding. Since Bernoulli arrivals have been assumed, this keeps the queue finite for $\lambda \leq 1$ and the optimal policy truncates the buffer up to length $L = l_{\pi^*}^{th}$. This reduces our investigation to the family of **L-truncating** policies (see Def. 2). The theorems provide furthermore an upper bound for the queue length. The importance of the penalty weight $\delta$ is emphasized by this expression. An increase in $\delta$ represents an increase of the dropping cost, which results in an increase of the optimal finite buffer length. In this case we prefer to increase system reliability and reduce the ratio of dropped packets at the cost of higher packet delay. Theorem 3 presents a condition regarding $\delta$ for which dropping is always optimal. Finally Theorems 1 and 4 prove the threshold behavior of the optimal policy on the k-axis. For $q_1 \geq q_2 \geq \ldots$ there exists a maximum positive integer $K = \max_l \hat{k}_l$ such that it is always optimal to drop for $k \geq K$ irrespective of $l$. This motivates the search over the optimal **K-retransmitting** policy (see Def. 1), which belongs to the family of policies for which retransmissions are allowed up to finite number of trials $K$. Observe by Theorem 4 that if $q_1 \leq q_2 \leq \ldots$ then $K$ is always unbounded (possibly constrained by the maximum queue length) since after some $l$-dependent threshold it is always optimal to continue.

Combining the two above suboptimal policies will provide a good approximation of the optimal strategy. Note that there will exist special values of $\delta$ and success probability sequences for which this is the optimal solution as well. The performance of the **L-truncating**, **K-retransmitting** policies, which will be from now on named $< \mathbf{L}, \mathbf{K} >$-policies, will be analyzed in the following paragraph.

## VI. Optimal and Suboptimal Delay-Dropping Tradeoffs

Since from Theorem 2 the maximum queue length and consequently the maximum number of retransmissions are always bounded, the state space is finite and standard algorithms such as policy iteration or value iteration can be implemented to determine the optimal solution. In the following, policy iteration for variable values of the dropping cost is used to provide the optimal delay and dropping tradeoff (fig.3 for decreasing respectively fig.6 for increasing success probabilities), as well as the behavior of the delay - expressed as average queue length $\bar{U}$ - (fig. 4 respectively fig. 7) and average dropping - $\lim_{N\to\infty} \frac{1}{N}\sum_{n=1}^{N} A_n p\left(Z_n\right)$ - (fig. 5 respectively fig. 8) with respect to the dropping cost $\delta$. The scenario implemented has arrival rate $\lambda = 0.6$ and $\beta = 0.99$. The decreasing sequence of success probabilities equals $q_k = e^{-0.3k}$ with $\max_k q_k = 0.741$, the increasing sequence of success probabilities equals $q_k = 1 - e^{-0.5k}$ with $\min_k q_k = 0.394$, maximum retransmission number $K_{max} = 10$ and queue length $L_{max} = 40$.

The results are compared to the $< L, K >$-policies suggested in the previous paragraph, where an algorithm similar to policy iteration but with certain adaptations is implemented to find the optimal L and K. The algorithm initializes with $\pi_0$ which is simply the policy which always choses drop as optimal action. At each step $n$ the policy $\pi_n := \pi_{<L_n,K_n>}$ is evaluated (see [18, pp. 174-175]) where $< L_n, K_n >$ are the max queue length and retransmission number and $J_{\pi_{<L_n,K_n>}}$ is obtained. There are four options. Either end the algorithm or look for a lower $J_{\pi_{n+1}}$ by increasing $K_n$, $L_n$ or both by 1. In this way comparing between these possibilities the algorithm evolves and terminates when $< L_n, K_n >=< L_{n+1}, K_{n+1} >$.

Observe that in both fig. 3 and fig. 6 the optimal delay-dropping tradeoff obtains a convex decreasing form. Lower delay implies necessarily a higher percentage of dropped packets. On the other hand the percentage of dropped packets for an arrival rate of $\lambda = 0.6$ and the given $\{q_k\}$ is only a rather small percentage of the transmitted packets ($< 3.2\%$ for decreasing success probabilities and $< 1.1\%$ for increasing). The delay (average dropping) increases (decreases) with respect to the dropping cost $\delta$, as fig. 4 and fig. 7 (fig. 5 and fig. 8) illustrate. Noteworthy is the fact that the maximum average number of dropped packets is much smaller in the case of increasing success probabilities compared to the decreasing case, although the first transmission effort has a rather low success probability $q_1 = 0.394$. Observing the plots it can be concluded that the $< L, K >$-policies have a near optimal behavior which was more or less expected since they are designed based on the optimal structural properties. Due to their simplicity they may be prefered to the actual optimal strategy since after determining the optimal retransmission number $K$, ARQ is applied for $l < L$, irrespective of the queue state, while for $l \geq L$ a one-shot transmission is made. Since the queue length cannot increase more than $L$ an interesting application could be, given $L$ as a design parameter for the maximum buffer length, to find the optimal maximum number of retransmissions of the ARQ chain.

## VII. Conclusions

We have considered in the current work a single queue which incorporates retransmissions of erroneous packets and can be dynamically controlled. The per effort success probabilities of the ARQ rounds are a priori defined after some possible offline optimization, since the model under study does not include channel state information at the transmitter. This is reasonable since the importance of ARQ lies in providing reliable communications simply with binary feedback. A Markov Decision Problem was formulated, where the action space is binary with possible actions (a) continue retransmission of erroneous packets and (b) drop the packet and receive a penalty $\delta$, while the cost per time slot is a linear combination of the current queue length and the penalty in case action (b) is chosen. Analysis of the structure of optimal policies $\pi^*$ has shown that there exists a queue length threshold $l^{th}_{\pi^*}$ after which retransmitting a packet is not any more optimal and the packets are sent in a single effort. Furthermore, given a queue length $l$, there exists a threshold $\hat{k}_l$ which suggests that if it is optimal to drop for retransmission number $\hat{k}_l$, the same is optimal $\forall k \geq \hat{k}_l$, in the case of monotone decreasing success probabilities. The monotonicity is inversed for increasing success probabilities. The results have motivated the investigation over structurally simpler delay-dropping policies which allow retransmissions as long as the retransmission number and the queue length remain below fixed thresholds $k \leq K$ and $l \leq L$, to be calculated. These are called here $< L, K >$-policies. The optimal delay-dropping tradeoff as well as the comparison of optimal and suboptimal strategies has been obtained using standard policy and value iteration algorithms and the results were illustrated in plots. These provide the possible average delay-dropping pairs to be obtained by appropriate choice of the penalty parameter $\delta$, depending on the demanded QoS. Furthermore the results support the near optimal behavior of the $< L, K >$-policies and suggest a simple, flexible, almost optimal alternative for simultaneously *reliable* and *delay-constrained* communications.

## APPENDIX

The following notation is introduced as a means to reduce space. For value iteration we write $TJ^n(l,k) = \min\left\{C_1^{n+1}(l,k), C_2^{n+1}(l,k)\right\}$, while the index $n$ is replaced by $\pi_n$ for the policy iteration steps. The dependence of the values $J_\beta^n(l,k)$ in value iteration and $J_{\pi_n,\beta}$ in policy iteration on $\beta$ is omitted.

$$\begin{aligned} C_1^{n+1}(l,k) &= l + \beta\left[q_k(1-\lambda) J^n(l-1,1) + q_k \lambda J^n(l,1) + \right. \\ &\quad \left. + p_k(1-\lambda) J^n(l,k+1) + p_k \lambda J^n(l+1,k+1)\right] \\ C_2^{n+1}(l,k) &= l + \delta p_k + \beta\left[\lambda J^n(l,1) + (1-\lambda) J^n(l-1,1)\right] \end{aligned}$$

Furthermore, rather useful for the analysis is the difference

$$\begin{aligned} \Delta J^{n+1}(l,k) = C_1^{n+1}(l,k) - C_2^{n+1}(l,k) = \\ -\delta p_k + \beta\left\{p_k(1-\lambda)\left[J^n(l,k+1) - J^n(l-1,1)\right] + p_k\lambda\left[J^n(l+1,k+1) - J^n(l,1)\right]\right\} \end{aligned} \tag{16}$$

### A. Monotonicity Properties

We always choose $J^0(l,k) = 0, \forall (l,k)$.

*Proof:* [Proof of Lemma 1] The inequality holds for $n = 0$. Suppose now that inequality (9) holds for iteration step $n$. We have to show that it also holds for $J^{n+1}(l,k) = TJ^n(l,k)$, the Dynamic Programming $T$ operator given in (8). We distinguish between two cases. (i) Suppose first that the 'continue' term $C_1^{n+1}(l+m,k)$ is the minimum. Then

$$\begin{aligned} TJ^n(l+m,k) \overset{(i)}{=} C_1^{n+1}(l+m,k) &\overset{(9)}{\geq} \\ C_1^{n+1}(l,k) \geq \min\left\{C_1^{n+1}(l,k), C_2^{n+1}(l,k)\right\} &= TJ^n(l,k) \end{aligned}$$

(ii) In the same fashion, we may reach the above inequality when the 'drop' term is the minimum, by replacing $C_1^{n+1}(l+m,k)$ by $C_2^{n+1}(l+m,k)$. Thus we have proved that in both cases $J^{n+1}(l,k) \leq J^{n+1}(l+m,k)$. ∎

*Proof:* [Proof of Lemma 2] The inequality certainly holds for $n = 0$. Suppose hypothesis (10) holds for some $n$. We proceed as in the proof of Lemma 1 and distinguish here the cases where the minimum in $TJ^n(l, k+m)$ is (i) $C_1^{n+1}(l, k+m)$ and (ii) $C_2^{n+1}(l, k+m)$. Then

$$TJ^n(l,k) \leq C_1^n(l,k) \overset{(\alpha)}{\leq} C_1^n(l,k+m) \overset{(i)}{=} TJ^n(l,k+m)$$

where $(\alpha)$ holds from hypothesis (10), $p_{k+m} \geq p_k$ and under the condition that $(1-\lambda)[J^n(l,k+1) - J^n(l-1,1)] + \lambda[J^n(l+1,k+1) - J^n(l,1)] \geq 0$, which is satisfied since $J^n(l,k+1) \overset{(9)}{\geq} J^n(l-1,k+1) \overset{(10)}{\geq} J^n(l-1,1)$. For the case (ii) the above inequality also holds replacing index 1 by 2 and $(\alpha)$ is simply due to (9) and the fact that $p_{k+m} \geq p_k$. Then we have proved that if $q_1 \geq q_2 \geq \dots$ inequality (10) holds $\forall n$. ■

## B. Proof of Theorem 1

*Proof:* Suppose continue is optimal for some state $\left(l, \hat{k}\right)$. Then using (16) we have $\Delta J^{n+1}\left(l, \hat{k}\right) \leq 0$. We have to prove that $\Delta J^{n+1}(l,k) \leq 0$, $\forall k < \hat{k}$. From (16) we have

$$\begin{aligned} &(1-\lambda) J^n\left(l, \hat{k}+1\right) + \lambda J^n\left(l+1, \hat{k}+1\right) \\ &\leq \delta/\beta + (1-\lambda) J^n(l-1,1) + \lambda J^n(l,1) \end{aligned}$$

Using Lemma 2 for the left handside we have $(1-\lambda) J^n(l,k+1) + \lambda J^n(l+1,k+1) \leq (1-\lambda) J^n\left(l, \hat{k}+1\right) + \lambda J^n\left(l+1, \hat{k}+1\right)$, $\forall k < \hat{k}$. Then there exists some maximum threshold value $\hat{k}_l \geq 1$ such that continue is optimal for $k < \hat{k}_l$ and drop for $k \geq \hat{k}_l$. The result holds $\forall n$ and consequently also for the optimal discounted and average reward policy as $n \to \infty$. ■

## C. Proof of Theorem 2

*Proof:* Consider a sequence $\pi_n,\ n = 1, 2, \dots$ of policies generated by the policy iteration algorithm which converge to $\pi^*$ as $n \to \infty$. Suppose that for a certain $n$ the policy $\pi_n$ has the following two properties

- **P1**: $\pi_n$ has a threshold value $l_{\pi_n}^{th}$ such that $\pi_n(l,k) = 1$, $\forall l \geq l_{\pi_n}^{th}$ & $\forall k$
- **P2**: The following inequalities hold $\forall l \geq l_{\pi_n}^{th}$

$$\begin{aligned} J_{\pi_n}(l+2,1) - J_{\pi_n}(l+1,1) &\geq J_{\pi_n}(l+1,1) - J_{\pi_n}(l,1) \\ J_{\pi_n}(l+1,1) &\geq J_{\pi_n}(l,1) \end{aligned}$$

(P2) implies that the difference $J_{\pi_n}(l,1) - J_{\pi_n}(l-1,1)$ is non-negative and monotone non-decreasing for the specified set of states.

We will prove that each policy $\pi_n$ as defined above, generates a $\pi_{n+1}$ with properties (P1) and (P2). Thus the optimal policy $\pi^*$ exhibits the same behavior.

We have to first provide a $\pi_0$ that satisfies (P1) and (P2) and initialize the policy iteration for $n = 0$. Consider the policy $\pi_0$ for which $\pi_0 = 1$, $\forall (l, k)$. In this case only a single transmission is allowed irrespective of the queue and retransmission state. The policy obviously has threshold $l^{th}_{\pi_0} = 0$ and property (P1) is fulfilled. We have to show that property (P2) also holds. This we will prove by induction. We first show the following inequality is satisfied for $l = 0$

$$J_{\pi_0}(2,1) - J_{\pi_0}(1,1) \geq J_{\pi_0}(1,1) - J_{\pi_0}(0,1)$$

Since always dropping takes place, using the expressions for $C_2(2,1), C_2(1,1), C_2(0,1)$

$$\begin{aligned} J_{\pi_0}(2,1) - J_{\pi_0}(1,1) = &\ 1 + \beta\lambda\left(J_{\pi_0}(2,1) - J_{\pi_0}(1,1)\right) + \beta(1-\lambda)\left(J_{\pi_0}(1,1) - J_{\pi_0}(0,1)\right) \\ J_{\pi_0}(1,1) - J_{\pi_0}(0,1) = &\ 1 + \delta p_1 \end{aligned}$$

Solving the first equation for $J_{\pi_0}(2,1) - J_{\pi_0}(1,1)$ and taking the difference of the above two

$$J_{\pi_0}(2,1) - 2J_{\pi_0}(1,1) + J_{\pi_0}(0,1) = \frac{\beta(1-\lambda)(1+\delta p_1) + 1}{1 - \beta\lambda} - 1 - \delta p_1$$

The above expression is positive if $\beta \geq 1 - \frac{1}{1+\delta p_1}$ hence the inequality holds for $\beta$ sufficiently close to 1, as $\delta$ ranges from 0 to $\infty$. Given now that the inequality in (P2) holds for $l - 1$ by induction we prove it also holds for $l$.

$$\begin{aligned} J_{\pi_0}(l+2,1) - J_{\pi_0}(l+1,1) \overset{(P1)}{=} \frac{1}{1-\beta\lambda}\left[1 + \beta(1-\lambda)\left(J_{\pi_0}(l+1,1) - J_{\pi_0}(l,1)\right)\right] \overset{l-1,(P2)}{\geq} \\ \frac{1}{1-\beta\lambda}\left[1 + \beta(1-\lambda)\left(J_{\pi_0}(l,1) - J_{\pi_0}(l-1,1)\right)\right] \overset{(P1)}{=} J_{\pi_n}(l+1,1) - J_{\pi_n}(l,1) \end{aligned}$$

Furthermore, since $J_{\pi_0}(l+1,1) - J_{\pi_0}(l,1) \geq J_{\pi_0}(1,1) - J_{\pi_0}(0,1) \geq 0$, the second inequality of (P2) is also proved to be true. Thus, both properties hold for $\pi_0$ which we can use to initialize the policy iteration algorithm.

We further continue using induction. Choose $l \geq l^{th}_{\pi_n} + 1$. Since we assume that $\pi_n$ satisfies (P1) and (P2), this implies that dropping occurs for the states $(l+1, k+1)$ as well as $(l, k+1)$.

$$\begin{aligned} J_{\pi_n}(l,k+1) - J_{\pi_n}(l-1,1) = C_2^{\pi_n}(l,k+1) - J_{\pi_n}(l-1,1) = \\ l + \delta p_{k+1} + \lambda\left[\beta J_{\pi_n}(l,1) - J_{\pi_n}(l-1,1)\right] + (1-\lambda)\left[\beta J_{\pi_n}(l-1,1) - J_{\pi_n}(l-1,1)\right] \end{aligned} \tag{17}$$

Observe now that for $\beta \to 1$, the last term vanishes. Since the difference $J_{\pi_n}(l,1) - J_{\pi_n}(l-1,1)$ is by (P2) non-decreasing for $l \geq l^{th}_{\pi_n} + 1$ then so is $J_{\pi_n}(l,k+1) - J_{\pi_n}(l-1,1)$. The same result holds for all queue states greater than $l$.

Let us now proceed to the Policy Improvement step of the Policy Iteration algorithm. Using the previous observation and the expression of $\Delta J_{\pi_{n+1}}(l,k)$ from (16)

$$\begin{aligned}\Delta J_{\pi_{n+1}}(l,k) &= -\delta p_k + \beta p_k (1-\lambda)\left[J_{\pi_n}(l,k+1) - J_{\pi_n}(l-1,1)\right] + \\ &+ \beta p_k \lambda \left[J_{\pi_n}(l+1,k+1) - J_{\pi_n}(l,1)\right] \end{aligned} \tag{18}$$

we conclude that $\Delta J_{\pi_{n+1}}(l,k)$ is non-decreasing w.r.t. $l$.

We cannot include the case $l = l_{\pi_n}^{th}$ in the above analysis since we do not know whether $J_{\pi_n}\left(l_{\pi_n}^{th}+1,1\right) - J_{\pi_n}\left(l_{\pi_n}^{th},1\right) \geq J_{\pi_n}\left(l_{\pi_n}^{th},1\right) - J_{\pi_n}\left(l_{\pi_n}^{th}-1,1\right)$. This we have first to prove. We know that for $\pi_n$ dropping occurs for $l = l_{\pi_n}^{th}$ and $l = l_{\pi_n}^{th}+1$. Then

$$\begin{aligned} J_{\pi_n}\left(l_{\pi_n}^{th}+1,1\right) - J_{\pi_n}\left(l_{\pi_n}^{th},1\right) &\overset{(P1)}{=} \tfrac{1}{1-\beta\lambda}\left[1+\beta(1-\lambda)\left(J_{\pi_n}\left(l_{\pi_n}^{th},1\right) - J_{\pi_n}\left(l_{\pi_n}^{th}-1,1\right)\right)\right] \\ \overset{\beta\to 1}{\Rightarrow} J_{\pi_n}\left(l_{\pi_n}^{th}+1,1\right) - J_{\pi_n}\left(l_{\pi_n}^{th},1\right) &= J_{\pi_n}\left(l_{\pi_n}^{th},1\right) - J_{\pi_n}\left(l_{\pi_n}^{th}-1,1\right) + \tfrac{1}{1-\lambda} \end{aligned} \tag{19}$$

and we conclude $\Delta J_{\pi_{n+1}}\left(l_{\pi_n}^{th}+1,k\right) \geq \Delta J_{\pi_{n+1}}\left(l_{\pi_n}^{th},k\right)$.

Then for each $k$ there exists a threshold $l_{\pi_{n+1}}^{th}(k)$ such that $\pi_{n+1}(l,k)=1$ for $l \geq l_{\pi_{n+1}}^{th}(k)$. Since the expression in (17) is non-decreasing and unbounded, the threshold for $\pi_{n+1}$ is always finite $\forall k$. We have $l_{\pi_{n+1}}^{th} = \max\left\{\max_k l_{\pi_{n+1}}^{th}(k), l_{\pi_n}^{th}\right\}$. The threshold defined in this way may either stay the same as in $\pi_n$ or increase (by a finite number of queue states). This we will use to verify (P2) for $\pi_{n+1}$.

Since the threshold satisfies $l_{\pi_{n+1}}^{th} \geq l_{\pi_n}^{th}$ then for all $(l,k)$ with $l \geq l_{\pi_{n+1}}^{th}$, dropping occurs according to policy $\pi_n$. Then $\forall l \geq l_{\pi_{n+1}}^{th}$ it holds

$$\begin{aligned} 2\cdot J_{\pi_{n+1}}(l+1,1) = 2\cdot C_2^{\pi_{n+1}}(l+1,1) &\overset{(P2)}{\leq} C_2^{\pi_{n+1}}(l+2,1) + C_2^{\pi_{n+1}}(l,1) \\ &= J_{\pi_{n+1}}(l+2,1) + J_{\pi_{n+1}}(l,1) \end{aligned}$$

and the inequality follows assuming that (P2) holds for $\pi_n$. For the second inequality we have for $l \geq l_{\pi_{n+1}}^{th}$, using the expression in (17) and $\beta \approx 1$

$$\begin{aligned} J_{\pi_{n+1}}(l+1,1) - J_{\pi_{n+1}}(l,1) \overset{(P1)}{=} C_2^{\pi_{n+1}}(l+1,1) - C_2^{\pi_{n+1}}(l,1) &\overset{(17)}{=} \\ 1 + \left[\lambda J_{\pi_n}(l+1,1) + (1-\lambda) J_{\pi_n}(l,1)\right] - \left[\lambda J_{\pi_n}(l,1) + (1-\lambda) J_{\pi_n}(l-1,1)\right] &\overset{(19)}{=} \\ J_{\pi_n}(l+1,1) - J_{\pi_n}(l,1) &\overset{(P2)}{\geq} 0 \end{aligned}$$

Hence policy $\pi_{n+1}$ shares the same properties as $\pi_n$ and the proof of the first part of the proposition is complete. For the second part using (18) together with (17) and $\beta \approx 1$ we can write after some manipulations

$$\Delta J_{\pi_{n+1}}\left(l_{\pi_n}^{th},k\right) = -\delta p_k + p_k\left(l_{\pi_n}^{th} + \delta p_{k+1}\right) + p_k\lambda\left[J_{\pi_n}\left(l_{\pi_n}^{th}+1,1\right) - J_{\pi_n}\left(l_{\pi_n}^{th},1\right)\right] - p_k(1-\lambda)$$

If $\Delta J_{\pi_{n+1}}\left(l_{\pi_n}^{th},k\right) \geq 0$ then the threshold will stay the same for $\pi_{n+1}$. This reduces to

$$l_{\pi_n}^{th}+\lambda\left[J_{\pi_n}\left(l_{\pi_n}^{th}+1,1\right)-J_{\pi_n}\left(l_{\pi_n}^{th},1\right)\right]\geq\delta q_{k+1}+(1-\lambda)$$

The left handside is an expression that depends only on $l_{\pi_n}^{th}$, whereas the right handside is a constant that depends on system parameters. The inequality will definitely be satisfied $\forall k$ for $l_{\pi_n}^{th}\geq\delta\max_k q_{k+1}+(1-\lambda)$, since by the second inequality of (P2) the difference in brackets is non-negative. Then using the policy iteration algorithm the aforementioned threshold cannot be exceeded and the threshold is always finite.

■

### D. Proof of Theorem 3

*Proof:* Assume that the policy iteration algorithm is initialized by $\pi_0$ as described in the proof of the previous theorem. The optimal policy will be $\pi^*=\pi_0$ if the threshold $l_{\pi_n}^{th}=0$, $\forall n$. Asssume that $\pi_n=\pi_0$. Let us first bound the difference $\Delta J_{\pi_{n+1}}(l,k)$. Since $J_{\pi_n}(l+1,1)-J_{\pi_n}(l,1)$ is increasing in $l$, $\forall l\geq 0$, then

$$\Delta J_{\pi_{n+1}}(l,k)\geq-\delta p_k+\beta p_k\lambda\left[J_n(2,k+1)-J_n(1,1)\right]+\beta p_k(1-\lambda)\left[J_n(1,k+1)-J_n(0,1)\right]$$

Omitting the details of the calculations, the following bound can be derived

$$\Delta J_{\pi_{n+1}}(l,k)\geq-\delta p_k+\beta p_k\left[1+\delta\left(p_{k+1}-p_1\right)\right]+\beta p_k\frac{\beta\lambda}{1-\beta\lambda}\left[1+\beta(1-\lambda)(1+\delta p_1)\right]$$

Then, if the right handside is $\geq 0\Rightarrow\Delta J_{\pi_{n+1}}(l,k)\geq 0$, $\forall(l,k)$. By simple calculations we get the expression in (13). The bound can be further written as $1/\left\{(1-\lambda)\left[1+(1-\lambda)p_1-\min_{k\neq 1}p_k\right]\right\}+\lambda/\left\{1+(1-\lambda)p_1-\min_{k\neq 1}p_k\right\}$, clearly non-decreasing w.r.t $\lambda$ and tending to $\infty$ for $\lambda\to 1$. ■

### E. Proof of Theorem 4

*Proof:* Let us show first that the two inequalities hold for policy $\pi_0$, where only dropping is chosen as action for all states. Inequality (15) is easy to verify since $J_{\pi_0}(l,k)=C_2(l,k)=l+\delta p_k+\beta\lambda J_{\pi_0}(l,1)+\beta(1-\lambda)J_{\pi_0}(l-1,1)\geq J_{\pi_0}(l,k+m)$, since $p_k\geq p_{k+m},\forall k\in\mathbf{Z}_+$. For the inequality (14) observe from the previous that $J_{\pi_0}(l,k)\geq J_{\pi_0}(l,k+m)$ holds for $m\to\infty$. We prove then that $J_{\pi_0}(l,k+m|_{m\to\infty})\geq J_{\pi_0}(l-1,1)$.

$$\begin{aligned}
J_{\pi_0}(l,k+m|_{m\to\infty})-J_{\pi_0}(l-1,1) &= \\
1-\delta p_1+\beta\lambda\left[J_{\pi_0}(l,1)-J_{\pi_0}(l-1,1)\right] &+ \\
\beta(1-\lambda)\left[J_{\pi_0}(l-1,1)-J_{\pi_0}(l-2,1)\right] &\overset{(a)}{\geq} \\
1-\delta p_1+\beta(1+\delta p_1) &\overset{(b)}{\geq} 0
\end{aligned}$$

where $(a)$ comes from Theorem 2, since the difference $J_{\pi_0}(l,1) - J_{\pi_0}(l-1,1)$ is monotone non-decreasing and $J_{\pi_0}(l,1) - J_{\pi_0}(l-1,1) \geq J_{\pi_0}(1,1) - J_{\pi_0}(0,1) = 1 + \delta p_1$ and $(b)$ holds for $\beta \geq 1 - 1/(1+\delta p_1)$ which tends to $1$ as $\delta \to \infty$.

Assume now that the policy iteration algorithm is initialized with policy $\pi_0$ and the above *two* inequalities hold for $\pi_n$. We will prove that the same holds for $\pi_{n+1}$, and hence the inequalities also hold for the optimal policy $\pi^*$ as $n \to \infty$. Let us first consider (14)

$$\begin{aligned} J_{\pi_{n+1}}(l,k) - J_{\pi_{n+1}}(l-1,1) &= \\ \min\{C_1^{\pi_{n+1}}(l,k), C_2^{\pi_{n+1}}(l,k)\} - J_{\pi_{n+1}}(l-1,1) &\overset{(c)}{\geq} \\ \min\{C_1^{\pi_{n+1}}(l,k), C_2^{\pi_{n+1}}(l,k)\} - J_{\pi_n}(l-1,1) &\overset{\beta\to 1,(d)}{\geq} 0 \end{aligned}$$

where $(c)$ follows from the property of the policy iteration algorithm that $\mathbf{J}_{\pi_{n+1}} \leq \mathbf{J}_{\pi_n}$ (for a proof the reader is referred to [18, Prop. 6.4.1, p. 175]) and $(d)$ comes from the induction argument and is easy to verify for $\beta \to 1$.

We continue to (15) and consider the two cases where (i) $d_{\pi_{n+1}}(l,k) = 1$, or (ii) $0$.

$$J_{\pi_{n+1}}(l,k) \overset{(i)}{=} C_2^{\pi_{n+1}}(l,k) \overset{(e)}{\geq} C_2^{\pi_{n+1}}(l,k+m) \geq J_{\pi_{n+1}}(l,k+m)$$

where $(e)$ is for the case (i) due to the fact that $p_k \geq p_{k+m}$. For case (ii) we have the same as above by changing the indices 2 with 1 and (i) with (ii). Inequality $(e)$ now follows from the fact that $p_k \geq p_{k+m}$ and $J_{\pi_n}(l,k) - J_{\pi_n}(l-1,1) \geq 0$ from the induction hypothesis.

For the *'Furthermore'* part of the Theorem, the proof follows the same lines as in that of Theorem 1 where inequality (15) is used in place of (10). ■

# TABLES

TABLE I

TRANSITION PROBABILITIES $\mathbf{Pr}\left(\bullet|S_n, A\right)$ OF THE MARKOV DECISION PROCESS UNDER STUDY

| $A_n = 0,\ U_n > 0$ | $\alpha_n = 0$ | $\alpha_n = 1$ |
|---|---|---|
| ACK | $\mathbf{Pr}\left(\left(U_n - 1, 1\right)\left|S_n, 0\right.\right) = q\left(Z_n\right)\left(1-\lambda\right)$ | $\mathbf{Pr}\left(\left(U_n, 1\right)\left|S_n, 0\right.\right) = q\left(Z_n\right)\lambda$ |
| NACK | $\mathbf{Pr}\left(\left(U_n, Z_n + 1\right)\left|S_n, 0\right.\right) = p\left(Z_n\right)\left(1-\lambda\right)$ | $\mathbf{Pr}\left(\left(U_n + 1, Z_n + 1\right)\left|S_n, 0\right.\right) = p\left(Z_n\right)\lambda$ |
| $A_n = 1,\ U_n > 0$ | $\mathbf{Pr}\left(\left(U_n - 1, 1\right)\left|S_n, 1\right.\right) = 1-\lambda$ | $\mathbf{Pr}\left(\left(U_n, 1\right)\left|S_n, 1\right.\right) = \lambda$ |
| $A_n = \{0, 1\},\ U_n = 0$ | $\mathbf{Pr}\left(\left(U_n, Z_n\right)\left|S_n, A_n\right.\right) = 1-\lambda$ | $\mathbf{Pr}\left(\left(U_n + 1, 1\right)\left|S_n, A_n\right.\right) = \lambda$ |

TABLE II

VALUE ITERATION FOR FINITE STATE SPACE

| **Example A - Bernoulli** | | | | | | | **Example B - Bernoulli** | | | | | | |
|---|---|---|---|---|---|---|---|---|---|---|---|---|---|
| L/K | **1** | **2** | **3** | **4** | **5** | **6** | L/M | **1** | **2** | **3** | **4** | **5** | **6** |
| **0** | 0 | 0 | 0 | 0 | 0 | 0 | **0** | 0 | 0 | 0 | 0 | 0 | 0 |
| **1** | 0 | 0 | 1 | 1 | 1 | 1 | **1** | 0 | 0 | 0 | 0 | 0 | 1 |
| **2** | 0 | 1 | 1 | 1 | 1 | 1 | **2** | 1 | 1 | 0 | 0 | 0 | 1 |
| **3** | 1 | 1 | 1 | 1 | 1 | 1 | **3** | 1 | 1 | 1 | 1 | 1 | 1 |
| **4** | 1 | 1 | 1 | 1 | 1 | 1 | **4** | 1 | 1 | 1 | 1 | 1 | 1 |
| **5** | 1 | 1 | 1 | 1 | 1 | 1 | **5** | 1 | 1 | 1 | 1 | 1 | 1 |
| **6** | 1 | 1 | 1 | 1 | 1 | 1 | **6** | 1 | 1 | 1 | 1 | 1 | 1 |
| **7** | 1 | 1 | 1 | 1 | 1 | 1 | **7** | 1 | 1 | 1 | 1 | 1 | 1 |
| **8** | 1 | 1 | 1 | 1 | 1 | 1 | **8** | 1 | 1 | 1 | 1 | 1 | 1 |
| **9** | 1 | 1 | 1 | 1 | 1 | 1 | **9** | 1 | 1 | 1 | 1 | 1 | 1 |
| **10** | 1 | 1 | 1 | 1 | 1 | 1 | **10** | 1 | 1 | 1 | 1 | 1 | 1 |

# FIGURES

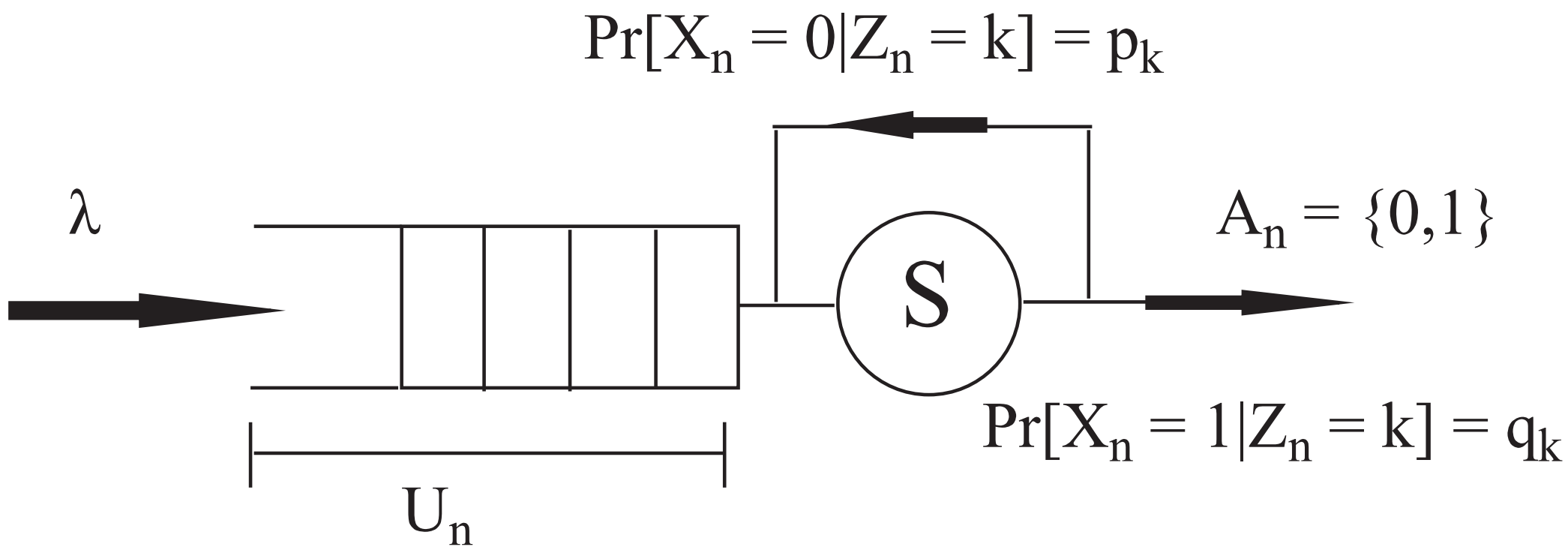

Fig. 1. A single server queue incorporating a retransmission protocol for the erroneous packets.

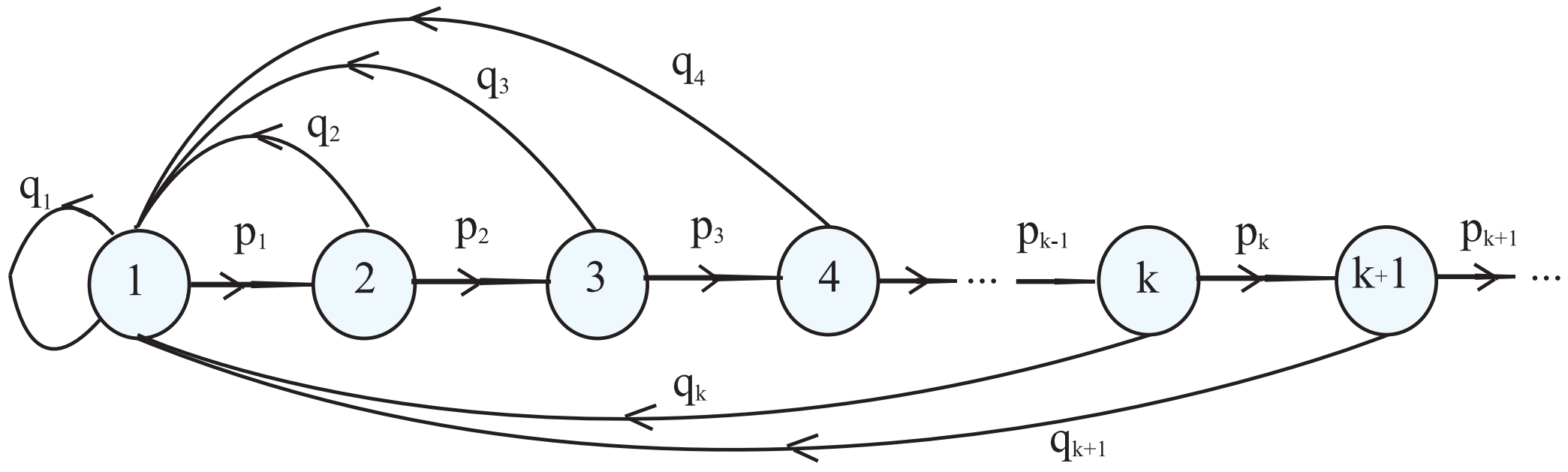

Fig. 2. Transition probability diagram for the random process $\{Z_n\}$ of current retransmissions. The ARQ Markov chain is considered time-homogeneous with countably infinite states.

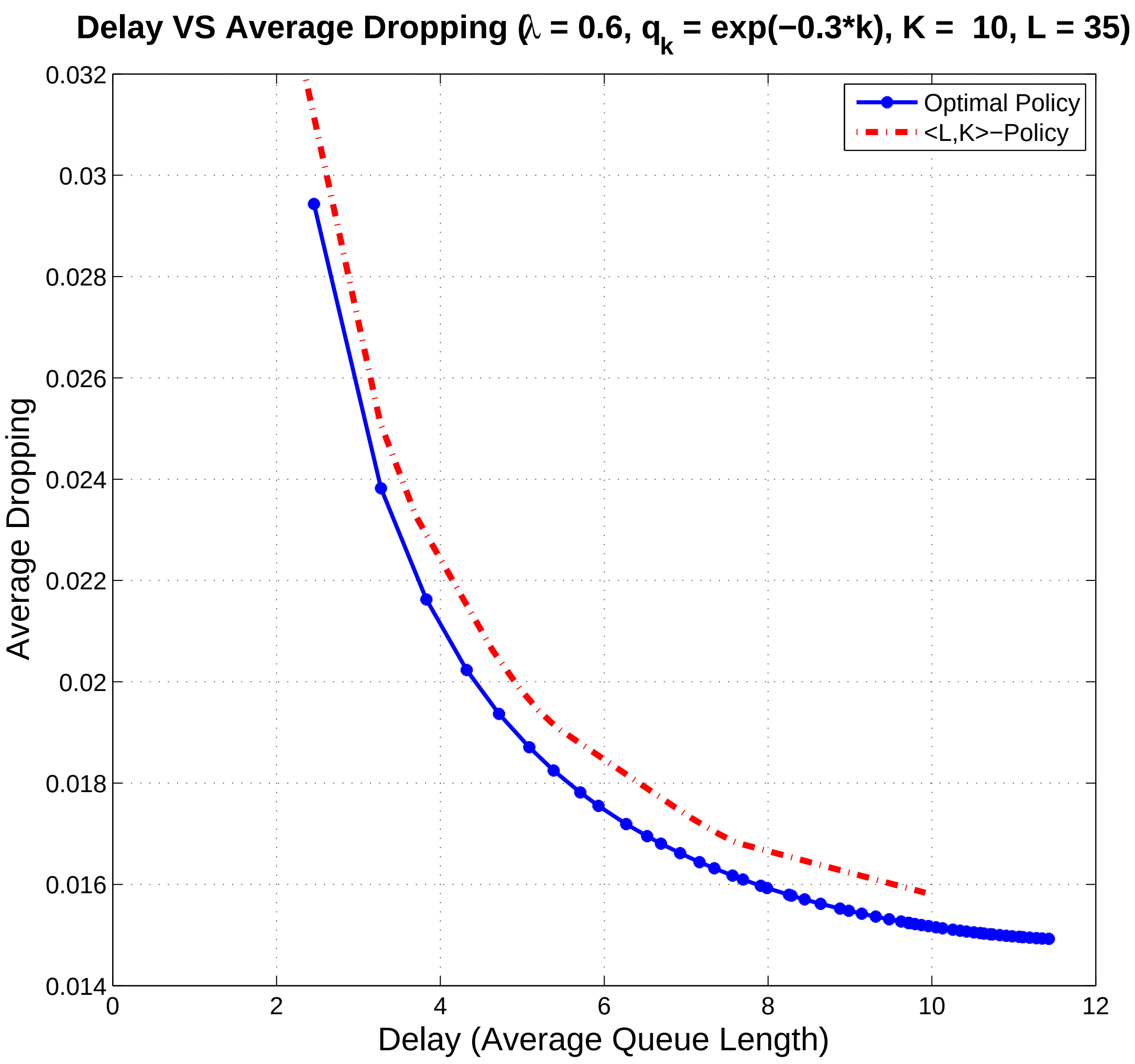

Fig. 3. The optimal and $< L, K >$-suboptimal delay-dropping tradeoff for decreasing sequence of success probabilities.

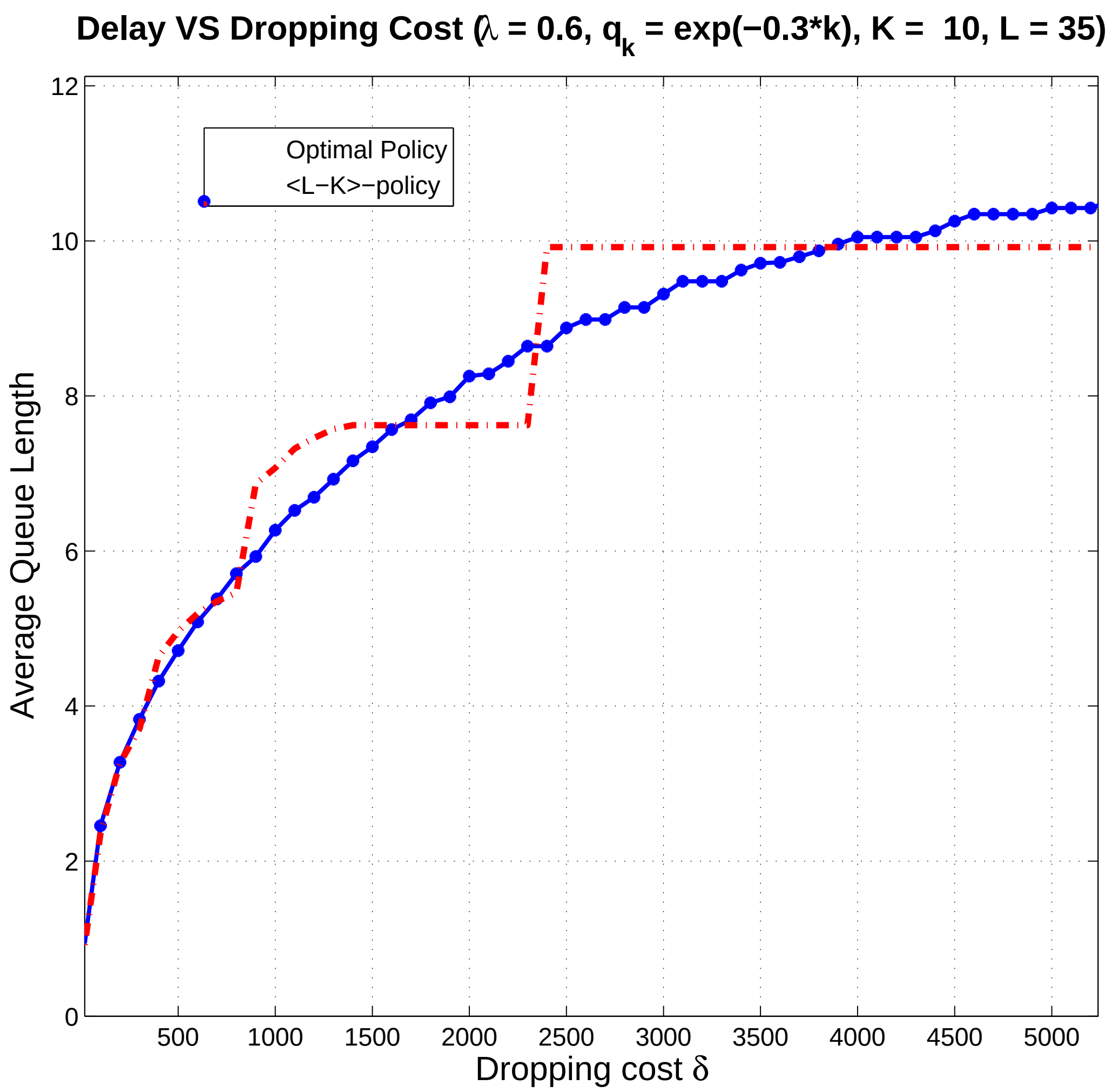

Fig. 4. Average queue length over increasing dropping cost for (a) the optimal policy and (b) the $< L, K >$-policies. The case of decreasing sequence of success probabilities is illustrated.

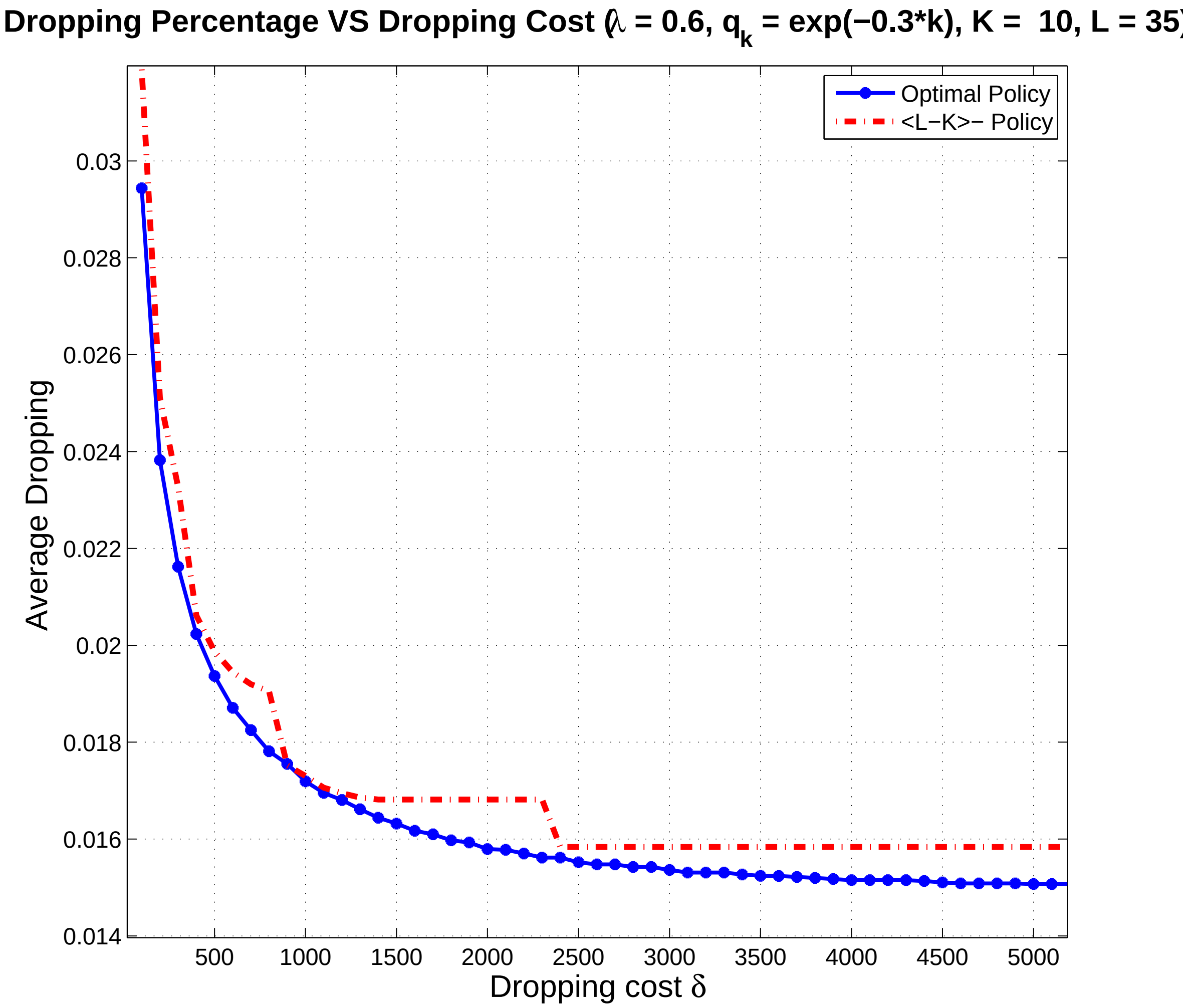

Fig. 5. Decrease of average dropping with increasing dropping cost for (a) the optimal policy and (b) the $< L, K >$-policies. The case of decreasing sequence of success probabilities is illustrated.

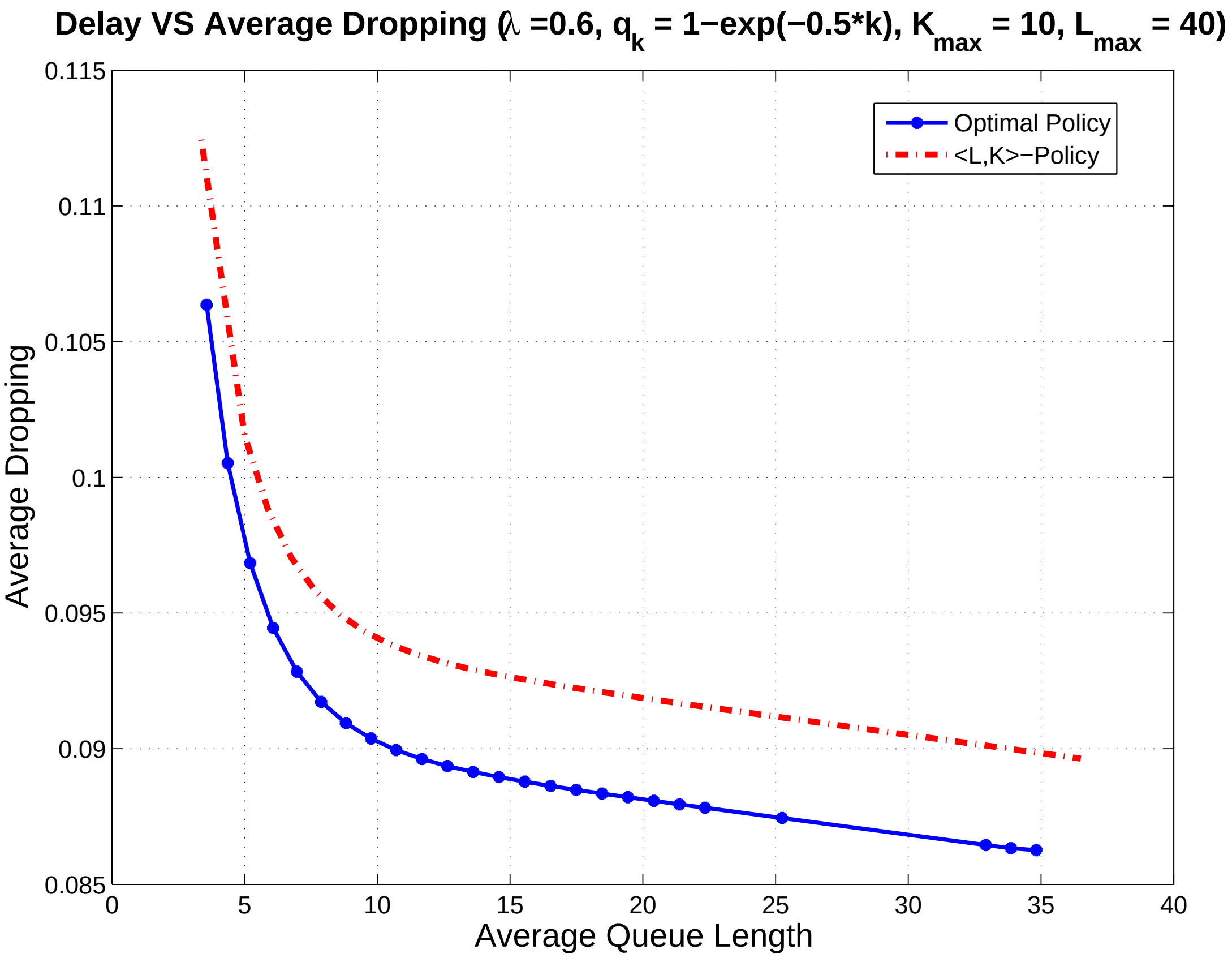

Fig. 6. The optimal and $< L, K >$-suboptimal delay-dropping tradeoff for increasing sequence of success probabilities.

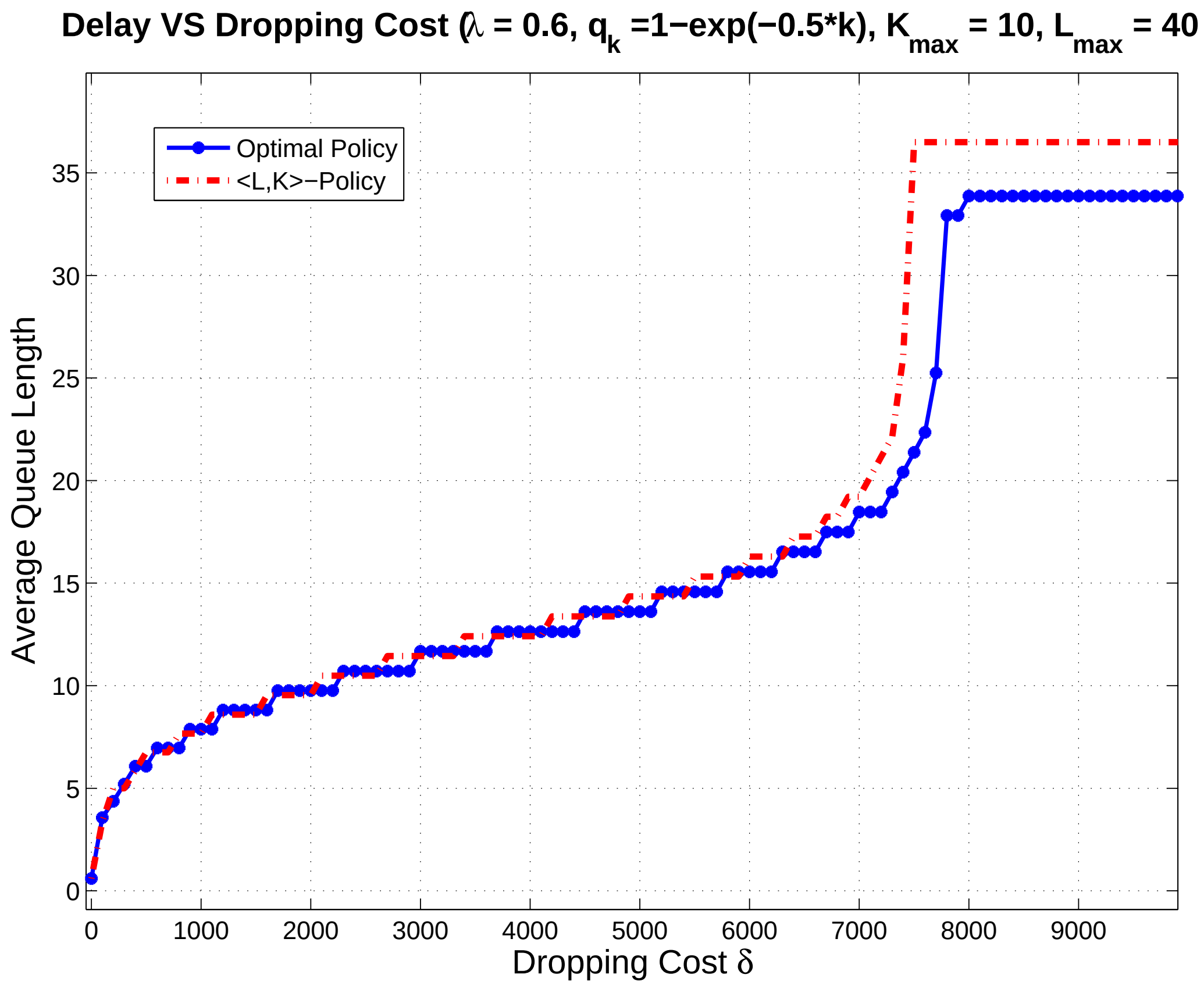

Fig. 7. Increase of average queue length with increasing dropping cost for (a) the optimal policy and (b) the $< L, K >$-policies. The case of increasing sequence of success probabilities is illustrated.

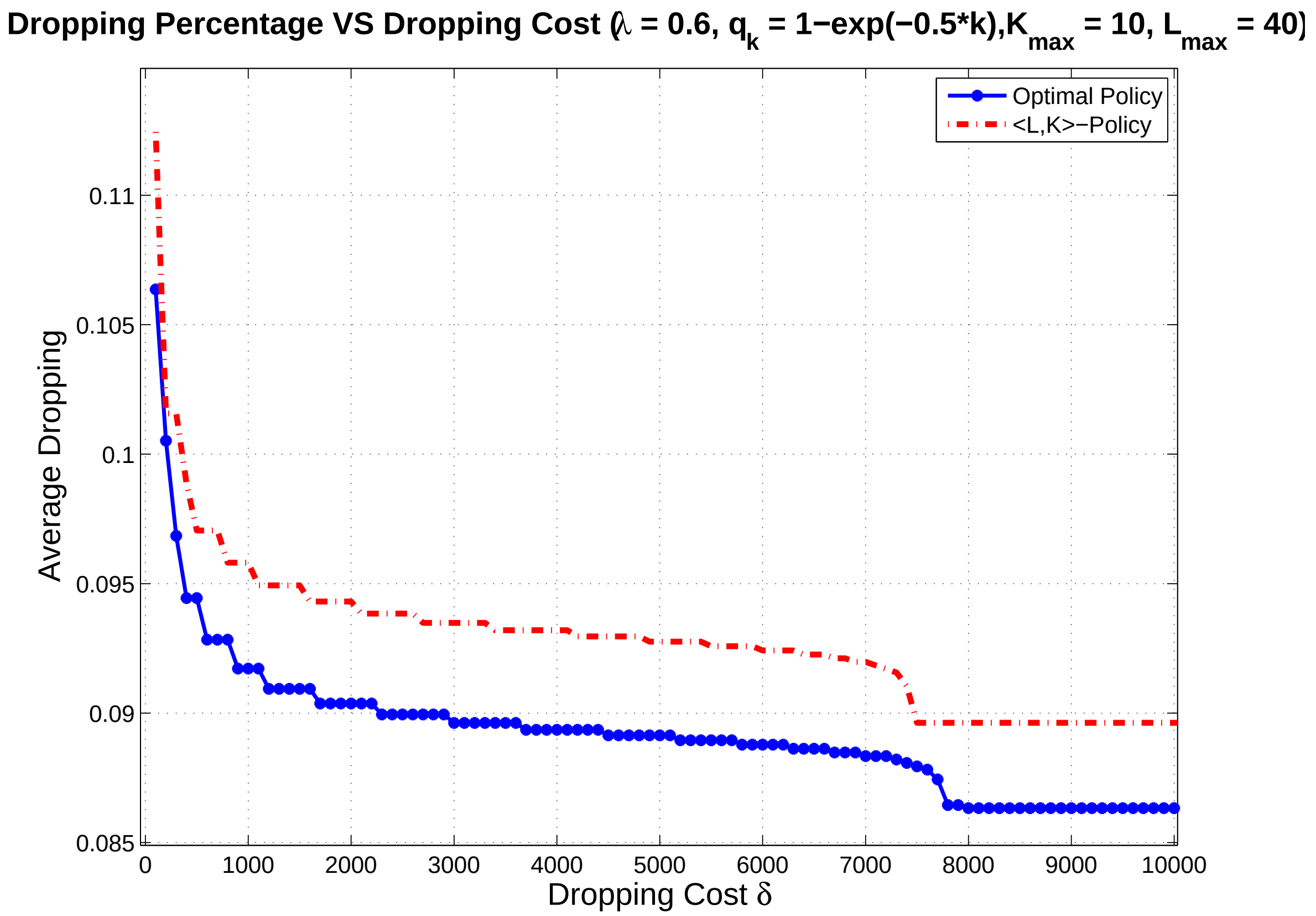

Fig. 8. Decrease of average dropping with increasing dropping cost for (a) the optimal policy and (b) the $< L, K >$-policies. The case of increasing sequence of success probabilities is illustrated.

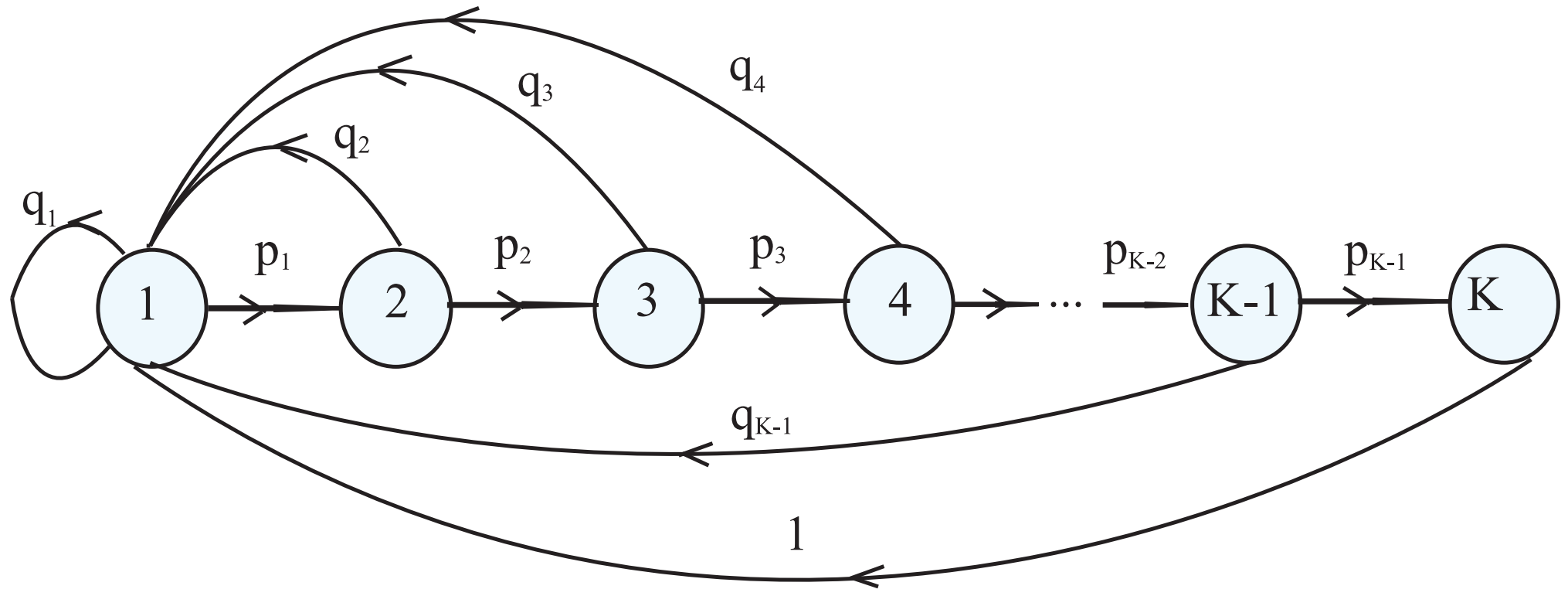

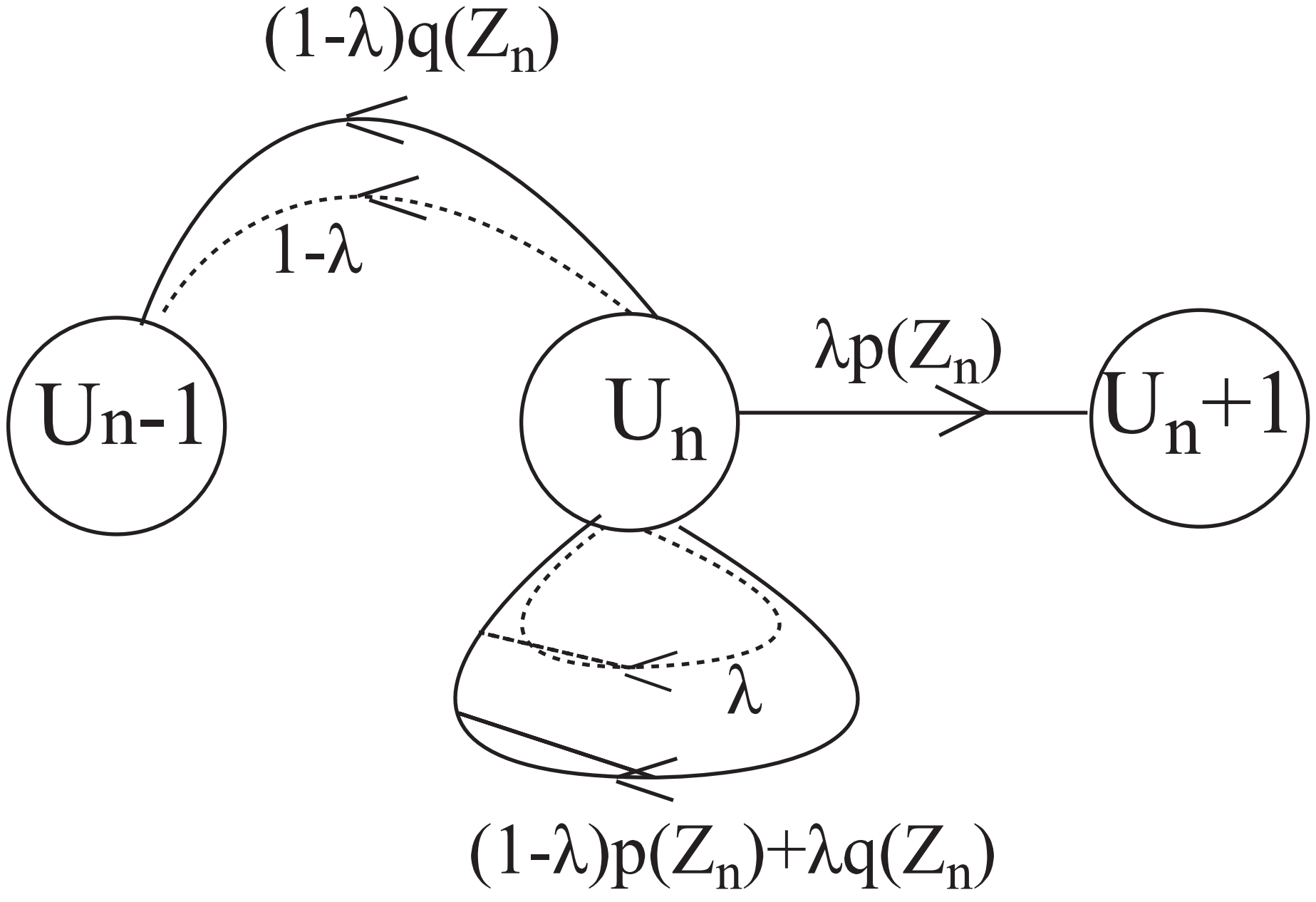

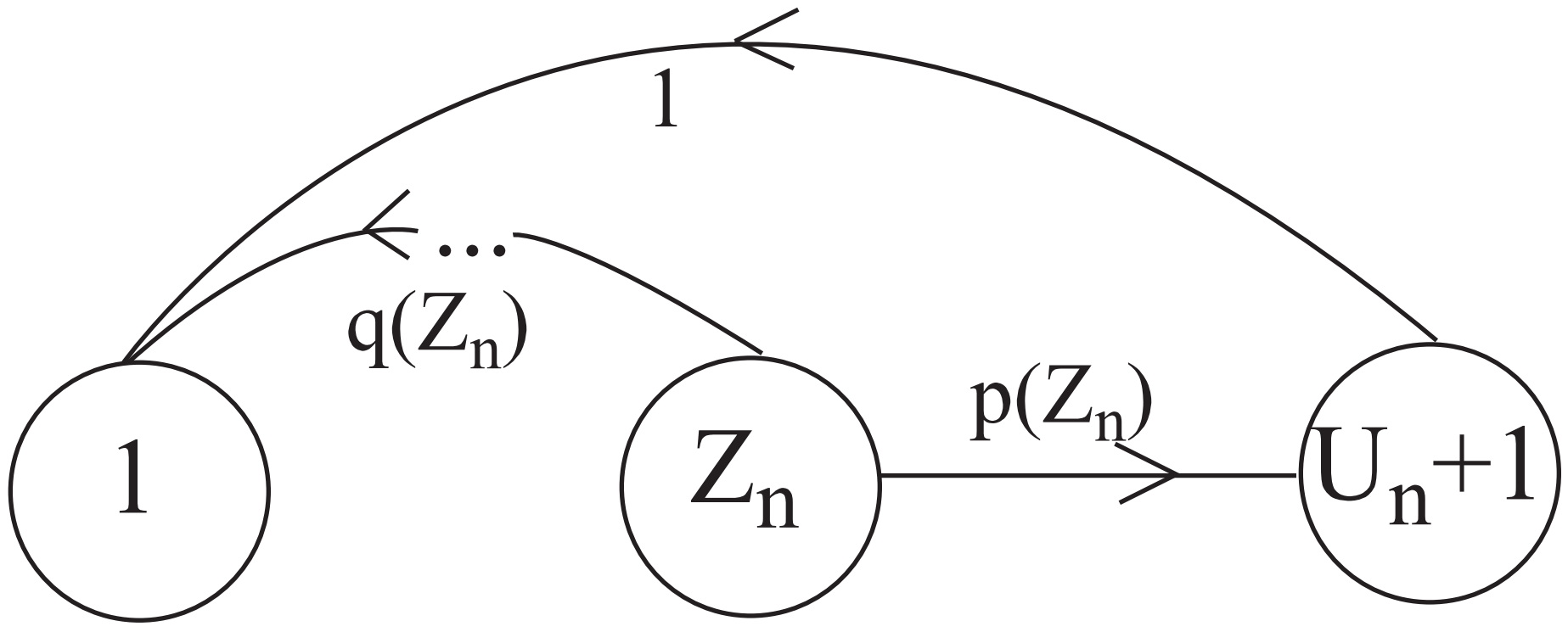